# Effect of Graphene on the Absorption and Extraordinary Transmission of Light in One Dimensional Metallic Gratings

Amirmasood Bagheri, Babak Rahmani, and Amin Khavasi

The authors are with the Electrical Engineering Department, Sharif University of Technology, Tehran 11155-4363, Iran

Email: khavasi@sharif.edu

*Abstract*
Periodic metallic structures are known to support resonant extraordinary transmission (EOT). When covered with graphene, these structures can be employed to effectively manipulate the light. In this work, we propose an analytical circuit model for graphene-covered one-dimensional metallic gratings. By using the circuit theory, we demonstrate that one-dimensional periodic array of cut-through slits which are covered by a continuous graphene sheet exhibit tunable EOT resonance whose amplitude, unlike its spectral position, can be dynamically tuned by varying the Fermi level of graphene. In this fashion, it is shown that placing a perfect reflector at the bottom of the graphene-covered metallic grating results in the realization of a graphene-based absorber. By utilizing the circuit theory, it is illustrated that perfect absorption in the structure is not exclusive to the TM polarization, but also the TE polarized plane waves can be completely absorbed by duly adjusting the Fermi level of graphene. Criteria for the enhanced absorption are accordingly presented. Results of this work may provide a useful tool for designing novel devices based on the graphene-covered metallic gratings such as filters, modulators and efficient absorbers.

## I. Introduction

Graphene, a two-dimensional material made of carbon atoms and arranged in honeycomb lattice, has exceptional electromagnetic properties including high electrical [1] and thermal conductivity [2], optical transparency [3], controllable plasmonic properties [4] and high operational speed [5]. We can think about new applications for graphene-based devices by harnessing these properties. Among such are transparent conductive electrodes [6], transformation optics [7], optical modulators [8] and absorbers [9].

Recently, the metal-graphene structures have spurred huge research interests. One-dimensional metallic gratings [10] and two-dimensional ring aperture-arrays [11] covered by a sheet of graphene are examples of structures with interesting optical features including tunable extraordinary transmission (EOT) and strong optical absorption. Thanks to the controllable plasmonic response in graphene, the former finds miscellaneous applications in filtering, optical modulation and switching [12]. The latter however, is used in devising light absorbers [13].

It is a well-known fact that the absorption of a monolayer graphene in the visible and near-infrared (NIR) regions is only about 2.3% [2]. Nevertheless, synergistic combination of graphene and metallic structures can substantially boost the absorption [13]. Additionally, the plasmonic response of graphene available in the mid- and far-infrared spectra can be also tailored to further enhance the absorption. Lately, it has been demonstrated that absorption of the incident light can be significantly increased via graphene-covered metallic gratings [14].

In addition to boosted absorption, graphene can be employed to achieve tunable and enhanced EOT. Combination of structured metallic gratings with graphene has proven a promising composite structure. Recently, it has been shown that coupling of graphene ribbons with an array of slits can render a transmission modulation of 95.7% in the mid-infrared spectrum [15]. Other hybrid structures such as graphene-gold nanoslit arrays make use of surface plasmon polaritons to increase mid-infrared EOT [16]. On the other hand, tuning of Fano-type transmission resonance via graphene in gratings has been reported in [17]. In a similar fashion, graphene-coated holey metal films have been propounded to provide tunable EOT resonances for sensing applications [18].

Since the discovery of EOT [19], metallic gratings have been under intensive research. Several numerical [20] and analytical approaches have been suggested for the analysis of EOT-supported structures [21]-[24]. Equivalent circuit models have been provided for two-dimensional arrays of holes [24] and patches [25], and one-dimensional symmetric [26] or asymmetric [27] arrays of slits. Nonetheless, there is a lack of circuit models for the accurate analysis of the graphene-covered metallic gratings. Recently, an equivalent structure based on the effective medium theory has been proposed [28]. That model, which can only be used for analyzing gratings with closed transmission channel, is incapable of predicting the behavior of higher diffracted orders in the periodic grating due to the its homogenized nature.

In this paper, we first propose an analytical circuit model for one-dimensional periodic arrangement of graphene-covered slits carved in a metallic film. To this end, the effect of higher diffraction orders necessary for the precise modeling of the grating as well as the plasmonic response of graphene are included in the model as a surface admittance created at the interface of the metallic grating and the surrounding media. Higher diffracted orders in the plain grating are produced due to the periodicity of the structure and can be addressed by an infinite number of parallel capacitors and inductors for the TM and TE polarizations, respectively. The effect of graphene sheet is also incorporated into the model as a circuit component. The proposed transmission line model is based on the single mode approximation within each slit and is valid for all incident angles. In order to validate the accuracy of the proposed model, several numerical examples are examined and the results are compared against those of the full-wave simulations performed by the commercial simulator COMSOL multiphysics.

Thanks to the proposed circuit model, tunability of EOT and enhancement of absorption in the graphene-covered metallic grating are elucidated in terms of parameters of the circuit model. It has been shown that the effect of graphene on the frequency of maximum absorption or EOT is negligible, although their amplitude can be substantially controlled by the graphene conductivity. We also show that perfect absorption is possible in TE polarization which has not been investigated in previous works.

The paper is organized as follows. Detailed description of the circuit model for one-dimensional periodic arrangement of slits is presented in section II. Section III is devoted to the simulation results and discussion. Conclusions are drawn in section IV. A time dependence of the form is assumed throughout this paper.

## II. Circuit Model

In this section, we propose an analytical circuit model for the analysis of one-dimensional periodic arrangement of slits carved in the PEC and covered by a continuous sheet of graphene. For simplicity's sake, we assume the grating is extended to infinity in the $z$-direction. Afterwards, we extend our model to a grating with a finite height. The schematic of the semi-infinite graphene-covered metallic grating is depicted in Fig. 1(a).

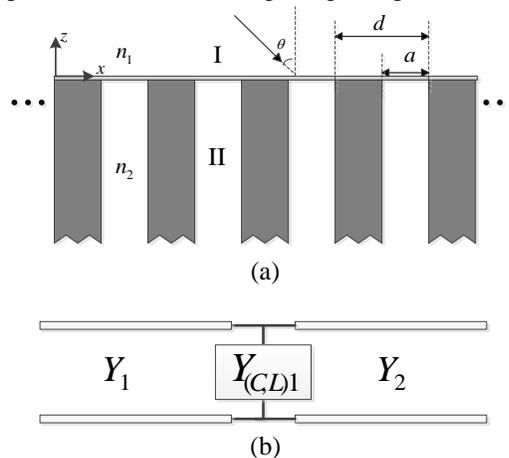

Fig. 1 a) Semi-infinite one-dimensional graphene-covered grating and b) its proposed circuit model

The aforementioned grating is periodic in the *x*-direction with period *d* and is uniform in the *y*-direction. The width of the slits is *a*. The refractive indices of the homogeneous medium above the grating (region I) and the slit region (region II) are $n_1$ and $n_2$, respectively. In order to model the graphene sheet, a surface conductivity modifying the continuity condition of tangential magnetic fields is being used. The conductivity of graphene is given by [9]:

$$\sigma_g = \frac{2e^2 k_B T}{\pi \hbar^2} \frac{j}{j\tau^{-1} - \omega} \ln\left[2\cosh\left(\frac{E_F}{2k_B T}\right)\right] \\ - \frac{je^2}{4\pi\hbar} \ln\left[\frac{2E_F - \hbar(\omega - j\tau^{-1})}{2E_F + \hbar(\omega - j\tau^{-1})}\right] \quad (1a)$$

where *e* is the electron charge, $E_F$ is the Fermi energy, $\hbar$ is the reduced Planck constant, $k_B$ is the Boltzmann constant, $\omega$ is the angular frequency, $T = 300$ K is the temperature and $\tau$ is the relaxation time. It should be noted, for sufficiently low frequencies and for $E_F \gg k_B T$, the graphene surface conductivity $\sigma_g$ is well described by the Drude form [9].

$$\sigma_g = \frac{e^2 E_F}{\pi \hbar^2} \frac{j}{j\tau^{-1} - \omega} \quad (1b)$$

In the rigorous full-wave calculations carried out by COMSOL Multiphysics, graphene is modeled by an ultrathin ($\Delta = 1$nm) layer whose permittivity is $\varepsilon_g = \varepsilon_0 - j\sigma_g/(\omega\Delta)$.

The proposed circuit model is derived for both major polarizations in the following subsections.

*A. TM Polarization*

Assuming that a TM polarized plane wave (the magnetic field in the y-direction) with incident angle $\theta$ with respect to the *z*-direction illuminates the structure, the tangential electric and magnetic fields in region I can be written in the following form:

$$H_{1y} = H_{10}^+ e^{jk_{z0} z} e^{-jk_{x0} x} + \sum_n H_{1n}^- e^{-jk_{zn} z} e^{-jk_{xn} x} \quad (2)$$

$$E_{1x} = -\xi_{10} H_{10}^+ e^{jk_{z0} z} e^{-jk_{x0} x} + \sum_n \xi_{1n} H_{1n}^- e^{-jk_{zn} z} e^{-jk_{xn} x} \quad (3)$$

where

$$k_{xn} = k_{x0} + \frac{2n\pi}{d}, \quad k_{x0} = k_0 n_1 \sin\theta, \quad n = 0, \pm 1, \pm 2,\ldots \quad (4a)$$

and

$$k_{zn} = k_0 \sqrt{n_1^2 - (n_1 \sin\theta + n\lambda/d)^2}, \\ \text{Re}[k_{zn}] > 0 \text{ or } \text{Im}[k_{zn}] < 0, \quad n = 0, \pm 1, \pm 2,.. \quad (4b)$$

are the wavevector components in the *x*- and *z*-directions, respectively. Also, $k_0$ is the wavenumber in the free space and $\lambda$ is the free space wavelength. Furthermore, $\xi_{1n} = k_{zn}/(\omega\varepsilon_0 n_1^2)$ is the wave impedance of the *n*th diffracted order in region I. It should be noted that, in a lossless medium, $k_{zn}$ can be either positive real or negative imaginary. The former corresponds to the propagating waves while the latter corresponds to the evanescent waves.

It should be noted that the width of the slits, i.e. *a*, is subwavelength enough to ensure only the principal TEM mode is propagating within each slit and all higher order modes are evanescent. The frequency range in which this approximation remains valid is $\lambda > 2an_2$ [21]. As a result, the electromagnetic fields in region II can be approximated by:

$$H_{2y} = H_{20}^+ e^{j\beta z} \tag{5}$$

$$E_{2x} = -\xi_{20} H_{20}^+ e^{j\beta z} \tag{6}$$

where

$$\xi_{20} = \frac{\beta}{\omega \varepsilon_0 n_2^2}, \quad \beta = k_0 n_2 \tag{7}$$

It should be noted that metals are assumed as perfect electric conductors (PECs) here. However, for real metals we should only modify the propagation constant. This modification leads to reasonably accurate results up to infrared regime [22], [29], [30]. Applying the continuity condition of tangential electric fields at the interface $z=0$ in a unit cell of the periodic structure yields

$$H_{10}^+ - H_{10}^- = H_{20}^+ \frac{\xi_{20}}{\xi_{10}} \sqrt{\frac{a}{d}} p_0^+ \tag{8}$$

$$H_{1n}^- = -H_{20}^+ \frac{\xi_{20}}{\xi_{1n}} \sqrt{\frac{a}{d}} p_n^+, \quad n \neq 0 \tag{9}$$

where

$$p_n^\pm = \frac{1}{\sqrt{ad}} \int_0^a e^{\pm j k_{xn} x} dx \tag{10}$$

These relations are obtained by multiplying the electric fields by $e^{j k_{xn} x}$ and then integrating both sides over the period $d$. On the other hand, we have

$$H_{10}^+ p_0^- + \sum_n H_{1n}^- p_n^- = H_{20}^+ \sqrt{\frac{a}{d}} [1 + \sigma_g \xi_{20}] \tag{11}$$

which is the boundary condition of tangential magnetic fields at the slit locations across the two regions obtained by including the effect of graphene as a surface conductivity. This relation is obtained by integrating both sides of the equation over the slit width. It should be noticed that metals are considered here as PECs, so the non-zero tangential electric field only occurs at the slit openings. Thereby, part of the graphene sheet which is placed on top of the metals has no effect on the boundary condition. By rearranging (8)-(11), one finds the following expression for the reflection coefficient:

$$\Gamma_0 = \frac{\frac{\xi_{20}}{\xi_{10}} |p_0^+|^2 - (1+\sigma_g \xi_{20}) - \sum_{n \neq 0} \frac{\xi_{20}}{\xi_{1n}} |p_n^+|^2}{\frac{\xi_{20}}{\xi_{10}} |p_0^+|^2 + (1+\sigma_g \xi_{20}) + \sum_{n \neq 0} \frac{\xi_{20}}{\xi_{1n}} |p_n^+|^2} \tag{12}$$

where $\Gamma_0 = -H_{10}^- / H_{10}^+$.

Remarkably, (12) resembles the reflection coefficient in the transmission line shown in Fig. 1(b):

$$\Gamma = \frac{Y_1 - [Y_2 + Y_{(C,L)1}]}{Y_1 + [Y_2 + Y_{(C,L)1}]} \tag{13}$$

Comparison of (12) and (13) suggests that an equivalent circuit model can be established for the graphene-covered metallic grating with $Y_1 = 1/\xi_{10}$ and $Y_2 = 1/(\xi_{20} |p_0^+|^2)$ being the corresponding characteristic admittances in the regions I and II, respectively. Furthermore, $Y_{(C,L)1}$ is given by

$$Y_{(C,L)1} = \frac{\sigma_g}{|p_0^+|^2} + \sum_{n \neq 0} \frac{|p_n^+|^2}{|p_0^+|^2 \xi_{1n}} \tag{14}$$

which accounts for the effects of higher order diffracted fields as well as graphene. As a matter of fact, $Y_{(C,L)1}$ consists of an infinite number of capacitors, i.e. $C_n = |p_n^+|^2 /(|p_0^+|^2 \xi_{1n})$, which are in parallel with the admittance attributed to the graphene conductivity. The admittance of graphene is inductive for $\hbar\omega < 5E_F/3$ [9]. When the $n$th diffracted order becomes propagating, its corresponding capacitor becomes resistive and can be substituted with a resistor whose value is $R_n = 1/(j\omega C_n)$. The power dissipated in $R_n$ is obtained using the circuit theory as follows:

$$P_{R_n} = \frac{1}{2}\frac{|V|^2}{R_n} \tag{15}$$

where $V$ is the voltage on the left transmission line and it can be written in terms of the incident wave amplitude and the reflection coefficient at $z=0$

$$V = V_0^+(1+\Gamma_0) \tag{16}$$

where $V_0^+$ is the incident wave amplitude. The incident power at $z=0$ also reads as

$$P_0 = \frac{1}{2}|V_0^+|^2 \operatorname{Re}(Y_1^*) \tag{17}$$

Therefore, the normalized power dissipated in $R_n$ can be written as

$$P_n = \frac{|(1+\Gamma_0)|^2}{\operatorname{Re}(Y_1^*) R_n} \tag{18}$$

As the surface admittance across the interface of the metallic grating and the homogeneous media are retrieved, we can extend our model to extract the appropriate equivalent circuit for the graphene-covered grating with finite height $h$. The structure and the proposed model are shown in Figs. 2(a) and 2(b), respectively. In the equivalent circuit, $Y_{(C,L)3}$ can be obtained using (14) and substituting $\xi_{1n}$ with $\xi_{3n}$ where $\xi_{3n} = k_{zn}/(\omega\varepsilon_0 n_3^2)$. It should be noted that $n_3$ is the refractive index of the homogeneous region below the grating. The characteristic admittance of region III is also given by $Y_3 = 1/\xi_{30}$. It should be emphasized that this structure is of great importance in the EOT context. The proposed circuit model can improve our understanding about how EOT is affected by graphene which is discussed in the next section.

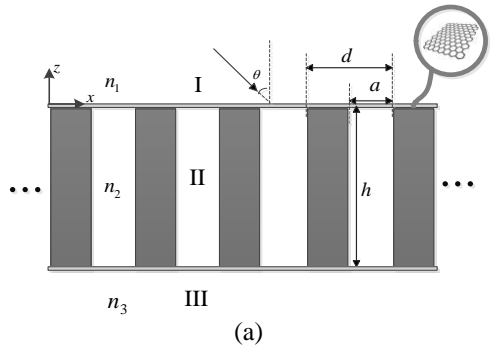

(a)

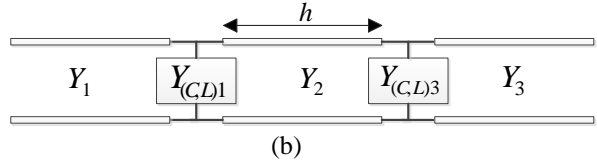

Fig. 2. a) The schematic of the one-dimensional graphene-covered metallic grating with finite height of $h$ and b) its equivalent circuit.

As in Fig. 3(a), placing a thick metallic slab at the bottom of the grating shorts $Y_{(C,L)3}$ and consequently the equivalent circuit model shown in Fig. 3(b) is obtained. This structure can be used as an absorber.

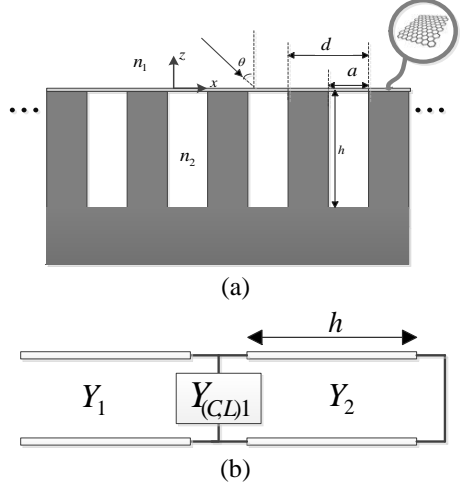

Fig. 3. a) The schematic of the one-dimensional graphene-covered metallic grating absorber and b) its equivalent circuit.

### B. TE Polarization

Similar to the procedure carried out for the TM polarization in the previous subsection, we presume the structure shown in Fig. 1(a) is illuminated by a TE polarized plane wave (the electric field in the $y$-direction) with incident angle $\theta$ with respect to the $z$-direction. The tangential electric and magnetic fields in region 1 can be written in the following form:

$$E_{1y} = E_{10}^{+} e^{jk_{z0} z} e^{-jk_{x0} x} + \sum_{n} E_{1n}^{-} e^{-jk_{zn} z} e^{-jk_{xn} x} \qquad (19)$$

$$H_{1x} = \varsigma_{10} E_{10}^{+} e^{jk_{z0} z} e^{-jk_{x0} x} - \sum_{n} \varsigma_{1n} E_{1n}^{-} e^{-jk_{zn} z} e^{-jk_{xn} x} \qquad (20)$$

where $\varsigma_{1n} = k_{zn}/\omega\mu$ is the wave admittance of the $n$th diffracted order in the region I. By assuming that only the principal TE mode is propagating inside the slits and all higher order modes are in cut-off, one can approximate the field distribution in the slits by:

$$E_{2y} = E_{20}^{+} \sin(\frac{\pi x}{a}) e^{j\beta z} \qquad (21)$$

$$H_{2x} = \varsigma_{20} E_{20}^{+} \sin(\frac{\pi x}{a}) e^{j\beta z} \qquad (22)$$

where

$$\varsigma_{20} = \frac{\beta}{\omega\mu}, \quad \beta = \sqrt{(k_0 n_2)^2 - (\pi/a)^2} \qquad (23)$$

The frequency range in which this approximation remains valid is $\lambda > a n_2$ [21]. Applying the continuity condition of tangential electric fields at the interface between the two regions, one can obtain:

$$E_{10}^+ + E_{10}^- = E_{20}^+ \sqrt{\frac{a}{2d}} q_0^+ \qquad (24)$$

$$E_{1n}^- = E_{20}^+ \sqrt{\frac{a}{2d}} q_n^+, \quad n \neq 0 \qquad (25)$$

where

$$q_n^\pm = \sqrt{\frac{2}{ad}} \int_0^a \sin(\frac{\pi x}{a}) e^{\pm j k_{xn} x} dx \qquad (26)$$

These relations are obtained by multiplying the electric fields by $e^{+jk_{xn}x}$ and then integrating both sides of the resultant equations over period $d$. Likewise, the boundary condition of the tangential magnetic field at slit openings across the interface $z=0$ reads as:

$$E_{10}^+ \varsigma_{10} q_0^- - \sum_n E_{1n}^- \varsigma_{1n} q_n^- = E_{20}^+ \sqrt{\frac{a}{2d}} [\sigma_g + \varsigma_{20}] \qquad (27)$$

where it is obtained by multiplying the magnetic fields by $\sin(\pi x/a)$ and integrating both sides over the slit width. By rearranging (23)-(27), one finds the following expression for the reflection coefficient:

$$\Gamma_0 = \frac{\frac{\varsigma_{10}}{\varsigma_{20}}|q_0^+|^2 - (1+\frac{\sigma_g}{\varsigma_{20}}) - \sum_{n \neq 0} \frac{\varsigma_{1n}}{\varsigma_{20}}|q_n^+|^2}{\frac{\varsigma_{10}}{\varsigma_{20}}|q_0^+|^2 + (1+\frac{\sigma_g}{\varsigma_{20}}) + \sum_{n \neq 0} \frac{\varsigma_{1n}}{\varsigma_{20}}|q_n^+|^2} \qquad (28)$$

where $\Gamma_0 = E_{10}^- / E_{10}^+$.

Similar to the TM case, (28) resembles the reflection coefficient of the transmission line in Fig. 1(b) with the parameters $Y_1 = \varsigma_{10}$, $Y_2 = \varsigma_{20}/|q_0^+|^2$ and:

$$Y_{(C,L)1} = \frac{\sigma_g}{|q_0^+|^2} + \sum_{n \neq 0} \frac{|q_n^+|^2}{|q_0^+|^2} \varsigma_{1n} \qquad (29)$$

It is noteworthy to mention that the shunt admittance between the homogeneous medium and the grating interface consists of an infinite number of inductors, i.e. $L_n = \varsigma_{1n}|q_n^+|^2/|q_0^+|^2$, which are parallel to the admittance associated with the graphene conductivity

### III. RESULTS AND DISCUSSION

In this section, several numerical examples are studied to verify the accuracy of the proposed model. Moreover, absorption enhancement of the graphene-covered metallic grating and tunable EOT are also discussed. It is noteworthy to mention that the rigorous full-wave simulations are performed with the commercial software COMSOL Multiphysics.

As the first example, we plot the zeroth-order reflected power from the structure depicted in Fig. 3(a) versus the normalized frequency $(d/\lambda)$ when the structure is illumined by a TM polarized plane wave. The geometrical parameters of the grating and the Fermi energy of graphene are $d = 200\mu m$, $n_1 = 1$, $n_2 = 1$, $a = 50\mu m$, $h = 100\mu m$ and $E_f = 0.1eV$, respectively. The relaxation time is chosen as $\tau = 2.5 \times 10^{-14}$ s hereafter. The accuracy of our circuit model is also studied for different incident angles. The results are shown in Fig. 4.

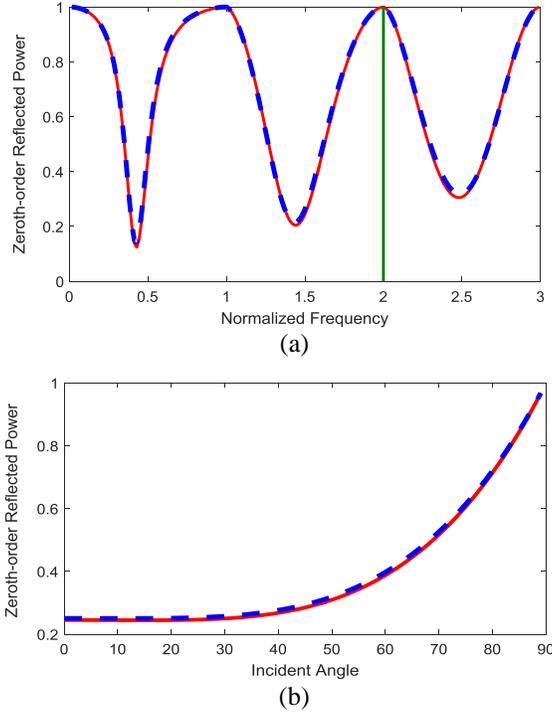

Fig. 4. a) Zeroth-order reflected power from a metallic grating covered by a graphene sheet (according to Fig. 3(a)) and under normal incidence versus the normalized frequency. b) Zeroth-order reflected power from the graphene-covered metallic grating in Fig. 3(a) versus the angle of incidence at the normalized frequency $d/\lambda = 0.39$. Results are calculated by the full-wave simulations (solid line) and the proposed method (dashed). The structure parameters are: $n_1 = 1, n_2 = 1$, $d = 200\mu m, h = 100\mu m$ and $a = 50\mu m$. The Fermi energy level is set to $E_f = 0.1eV$. The straight solid line in (a) shows the frequency range within which the single mode approximation condition holds.

As shown in Figs. 4, there is an excellent agreement between the results obtained by the full-wave simulations and those of the proposed circuit model, even for the normalized frequencies beyond the single mode limit. This is due to the fact that under normal incidence, the impinging wave has an even symmetry so it cannot be coupled to the higher order mode in the parallel plate waveguide formed by the PEC walls of the slits which has an odd symmetry.

The normalized power transmitted to the ±1 st-reflected order in the previous example is also plotted versus the normalized frequency in Fig. 5. Evidently, results of the full-wave simulations and those of the proposed circuit model are consistent with each other. As it can be inferred from Fig. 5, when the ±1 st $(d/\lambda \geq 1)$ diffraction orders become propagating, their corresponding capacitors $(C_{\pm 1})$ become resistive $(R_{\pm 1})$.

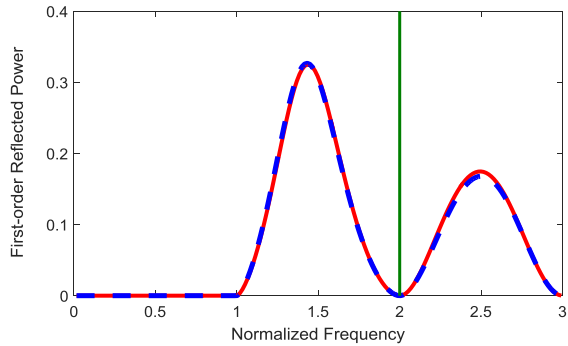

Fig. 5. The normalized power transmitted to the ±1 st reflected order in the graphene-covered metallic grating of Fig. 3(a) with parameters $n_1 = 1$, $n_2 = 1$, $d = 200\mu m$, $h = 100\mu m$ and $a = 50\mu m$ and under normal incidence versus the normalized frequency. Results are calculated by the full-wave simulations (solid line) and the proposed method (dashed). The Fermi energy level is set to $E_f = 0.1eV$. The straight solid line shows the frequency range within which the single mode approximation condition holds.

In order to further demonstrate the usefulness of the proposed model, we investigate the enhanced absorption of graphene-covered metallic grating of Fig. 3(a) using our model. Figure 6 depicts the total absorption of a normally incident TM wave versus the normalized frequency for different values of Fermi energy level in the previous example.

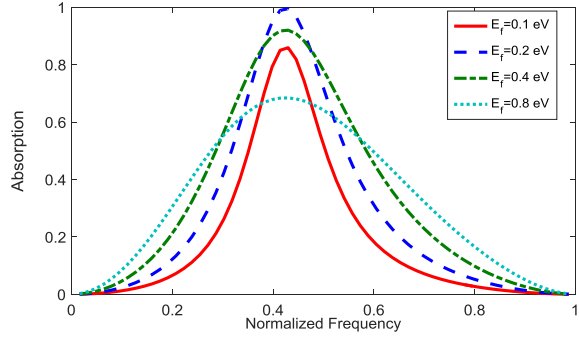

Fig 6. Total absorption in the graphene-covered metallic grating of Fig. 3(a) with parameters $n_1 = 1$, $n_2 = 1$, $d = 200\mu\text{m}$, $h = 100\mu\text{m}$ and $a = 50\mu\text{m}$ for different Fermi energy levels versus the normalized frequency when illuminated by a TM polarized normally incident plane wave.

It is obvious from Fig. 6 that for different values of Fermi energy level, the location of the peak in the absorption spectrum is almost constant. On the other hand, the amount of absorption strongly depends on the Fermi energy. These features can be better understood by resorting to the equivalent circuit model of Fig. 3(b). Using the transmission line theory, the reflection coefficient in this circuit is

$$\Gamma = \frac{Y_1 - [Y_{(C,L)1} - jY_2\cot(\beta h)]}{Y_1 + [Y_{(C,L)1} - jY_2\cot(\beta h)]} \quad (30)$$

Evidently, the condition for perfect absorption (which is equivalent to zero reflection due to the closed transmission channel in this structure) is $Y_1 = Y_{(C,L)1} - jY_2\cot(\beta h)$, which translates into the impedance matching at the interface of the homogeneous medium above the grating. For the normalized frequencies less than unity, this condition is equivalent to

$$Y_1 = \frac{\text{Re}[\sigma_g]}{|p_0^+|^2} \quad (31)$$

and the following resonance condition:

$$Y_2\cot(\beta h) - \frac{\text{Im}[\sigma_g]}{|p_0^+|^2} - \sum_{n \neq 0}\text{Im}[C_n] = 0 \quad (32)$$

We note that the second term in (32) is orders of magnitudes smaller than the third term and thus it can be neglected. Consequently, the resonance frequency of the enhanced absorption is almost independent of the graphene conductivity. Additionally, it is expected that the strong absorption occur when the Fabry-perot resonance condition $2\beta h \approx \pi$ is satisfied. Nonetheless, surface capacitors at the interface of the metallic grating and the homogenous medium slightly change the resonance frequency. The resonant frequency $\omega_r$ can be obtained by solving the transcendental equation of (32). On the other hand, one can easily obtain the Fermi energy level of graphene required for the realization of a perfect absorber from (31) calculated at $\omega = \omega_r$:

$$E_f = |p_0^+|^2 Y_1 \frac{\pi\hbar^2(\omega^2 + \tau^{-2})}{e^2\tau^{-1}} \quad (33)$$

wherein, the Drude-like conductivity of graphene according to (1b) is evoked. For instance, we obtain the optimum value of Fermi energy in the example of Fig. 6 as $E_f = 0.225\text{eV}$ which is in good agreement with simulation results.

As the final example for the TM polarization, the transmission of the incident light through the structure depicted in Fig. 2(a) is investigated by plotting the transmitted power versus the normalized frequency for different Fermi energy levels. The geometrical parameters of the grating are $d = 200\mu m$, $n_1 = 1$, $n_2 = 1$, $a = 50\mu m$, $h = 100\mu m$. The results are shown in Fig. 7. It can be inferred from Fig. 7 that the peak positions of EOTs are almost constant, but their amplitude greatly depends on the Fermi energy level of graphene and thus can be easily tuned.

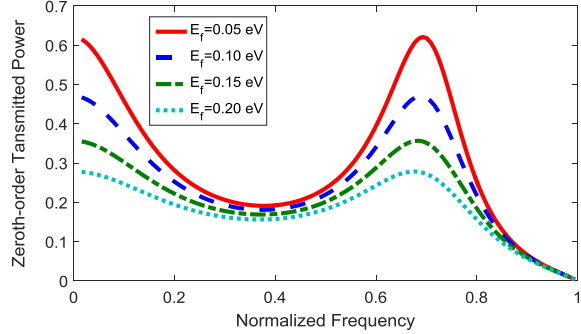

Fig. 7. Transmitted power from the graphene-covered metallic grating of Fig. 2(a) with parameters $n_1 = 1$, $n_2 = 1$, $d = 200\mu m$, $h = 100\mu m$ and $a = 50\mu m$ for different Fermi energy levels $E_f = 0.05 \text{eV}$ (solid line), $E_f = 0.1 \text{eV}$ (dashed line), $E_f = 0.15 \text{eV}$ (dashed-dotted line) and $E_f = 0.2 \text{eV}$ (dotted line) versus the normalized frequency.

Next, we examine the validity of our model for the TE polarization. The zeroth-order reflected power from the graphene-covered metallic grating depicted in Fig. 3(a) with parameters $n_1 = 1$, $n_2 = 1$, $d = 200\mu m$, $h = 200\mu m$ and $a = 110\mu m$, $E_f = 0.2 \text{eV}$ and under the normal incidence is plotted versus the normalized frequency in Fig. 8(a). The accuracy of the circuit model is also investigated for different incident angles. The results are shown in Fig. 8(b). For the normalized frequencies below the single mode approximation limit, excellent agreement is observed between the results of the rigorous method and those of the proposed circuit model.

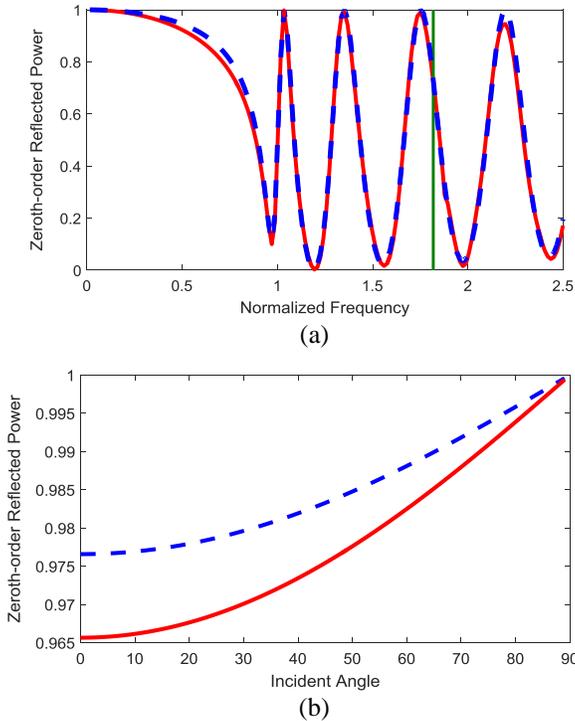

Fig. 8. a) Zeroth-order reflected power from a metallic grating covered by a graphene sheet (according to Fig. 3(a)) and under normal incidence versus the normalized frequency. b) Zeroth-order reflected power from the graphene-covered metallic grating in Fig. 3(a) versus the angle of incidence at the normalized frequency $d / \lambda = 0.35$. Results are calculated by the full-wave simulations (solid line) and the proposed method

(dashed). The structure parameters are: $n_1 = 1$, $n_2 = 1$, $d = 200\mu m$, $h = 200\mu m$ and $a = 110\mu m$. The Fermi energy level is considered to be $E_f = 0.2\,eV$. The straight solid line in (a) shows the frequency range within which the single mode approximation condition holds.

As the final example for the TE case, the enhanced absorption of graphene-covered metallic grating of Fig. 3(a) is studied using our model. Fig.9 depicts the total absorption of a normally incident TE wave versus the normalized frequency for different values of Fermi energy level in the previous example. As evident from Fig. 9, enhanced absorption in the graphene-covered metallic grating when illuminated by a TE polarized plane wave is achieved. It is in sheer contrast to the previous works [13] which excluded the TE polarization from achieving perfect absorption in the one-dimensional graphene-covered metallic gratings. In parallel with the TM case, the design criteria for perfect absorption in TE polarization can be easily obtained. For the normalized frequencies less than unity, the condition for perfect absorption is derived by substituting $\left|p_0^+\right|^2$ and $C_n$ in Eqs. (30) and (31) with $\left|q_0^+\right|^2$ and $L_n$, respectively. Moreover, substitution of $\left|p_0^+\right|^2$ in Eq. (32) with $\left|q_0^+\right|^2$ gives the optimum value of Fermi energy of graphene for the perfect absorption ($E_f = 0.400\,eV$ in this case).

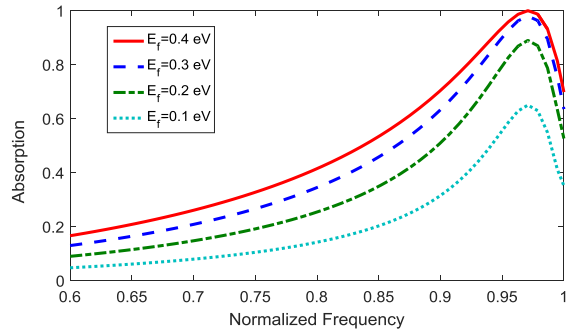

Fig 9. Total absorption in the graphene-covered metallic grating of Fig. 3(a) with parameters $n_1 = 1$, $n_2 = 1$, $d = 200\mu m$, $h = 200\mu m$ and $a = 110\mu m$ for different Fermi energy levels versus the normalized frequency when illuminated by a TE polarized normally incident plane wave.

## IV. CONCLUSION

In this work, we proposed an equivalent circuit model for the analysis of one-dimensional graphene-covered metallic gratings by using the transmission line theory. The parameters of the proposed circuit model are given by explicit analytical expressions for both major polarizations and for any arbitrary incident angle. It is demonstrated that the effect of higher diffracted orders in the periodic grating as well as the plasmonic response of graphene can be well described by introducing a surface admittance across the interface of the grating and its immediate surroundings.

The accuracy of the model is verified by the full-wave numerical simulations. It is shown that our model is capable of predicting the behavior of the original structure provided that only the principal mode is propagating inside the slits.

Using this model, we showed the absorptive nature of the structure can be well described by using the proposed equivalent circuit model. Remarkably, we illustrated that perfect absorption can be also realized for the TE polarization. Moreover, criteria for designing perfect absorbers have been discussed. We demonstrated that the peak positions of both the EOT and absorption spectra remain almost constant for different values of the Fermi energy level of graphene. However, their amplitudes can be significantly tuned by changing the Fermi level. This work provides a powerful tool for design and analysis of plasmonic devices based on the graphene-covered metallic gratings.